\documentclass[]{aa}       
\usepackage{graphicx}
\usepackage{amssymb}
\usepackage{amsmath}
\usepackage{longtable}
\usepackage{supertabular}
\usepackage{natbib} 

\def\fer{{\it Fermi }}

\def\be{\begin{equation}}
\def\ee{\end{equation}}
\def\ba{\begin{eqnarray}}
\def\ea{\end{eqnarray}}

\def\v{\upsilon}

\def\mkn501{\object{Mkn~501}}

\def\fb{{\it Fermi} bubbles}

\begin{document}

\title{Formation of large-scale magnetic structures associated with the \fer{} bubbles}

\author{M. V. Barkov \inst{1,2} \and 
        V. Bosch-Ramon \inst{3}         
    }

\authorrunning{Barkov and Bosch-Ramon}

\titlerunning{Formation of magnetic-field structure in the \fer{} bubbles}

\institute{Max-Planck-Institut f\"ur Kernphysik, Saupfercheckweg 1, 69117 Heidelberg, Germany; bmv@mpi-hd.mpg.de
\and
Space Research Institute RAS, 84/32 Profsoyuznaya Street, Moscow, 117997, Russia
\and
Departament d'Astronomia i Meteorologia, Institut de Ci\`encies del Cosmos (ICC), Universitat de Barcelona (IEEC-UB), Mart\'i i
Franqu\`es 1, 08028 Barcelona, Spain, e-mail: vbosch@am.ub.es
}

\offprints{M. V. Barkov, \email{bmv@mpi-hd.mpg.de}}

\date{Received <date> / Accepted <date>}

\abstract
{The {\it Fermi}{} bubbles are part of a complex region of the Milky Way. This region presents broadband extended non-thermal 
radiation, apparently coming
from a physical structure rooted in the Galactic Centre and with a { partly-ordered magnetic field}{} threading it.}  
{We explore the possibility of an explosive origin for the {\it Fermi}{} bubble region to explain its morphology, in particular that 
of the large-scale magnetic fields, and provide context for the broadband non-thermal radiation.} 
{We perform 3D magnetohydrodynamical simulations of an explosion from a few million years ago that pushed and sheared a surrounding
magnetic loop, anchored in the molecular torus around the Galactic Centre.}
{Our results can explain the formation of the large-scale magnetic structure in the {\it Fermi}{} bubble region. Consecutive explosive events may
match better the morphology of the  region. Faster velocities at the top of the shocks than at their sides
may explain the hardening with distance from the Galactic Plane found in the GeV emission.}
{In the framework of our scenario, we estimate the lifetime of the {\it Fermi}{} bubbles as $\approx 2\times10^6$~yr, with a total energy
injected  in the explosion(s) $\gtrsim 10^{55}$~ergs. The broadband non-thermal radiation from the region may be explained by leptonic
emission, more extended in radio and X-rays, and confined to the {\fb{}} in gamma rays.} 
\keywords{Shock waves, Methods: numerical, Galaxy: center }

\maketitle

\section{Introduction} \label{intro}

{Originating in the Galactic Centre and extending up to a distance
of $\approx 15$~kpc from the Galactic plane, there is a large bipolar structure with counterparts in different wavelengths, from radio to
gamma rays.} This structure was seen for the first time 
by {\it ROSAT} in X-rays \citep{sef97}. An overlapping extended source was also detected
in radio by WMAP \citep{fin04}, and a few years later \fer{} detected a somewhat smaller structure in the
GeV range, the \fb{}, with a height of $\approx 8$~kpc \citep{ssf10}. A hint of an extension of the GeV
emission beyond the \fb{} was also been found on scales similar to those of the ROSAT emission \citep{fermi12}.
{S-PASS \citep{CCS13} and WMAP \citep{jcro12} detected a large-scale, partly-ordered magnetic-field structure placed at $20^o-50^o$
from the galactic plane in the \fer{} bubble region, but larger than the bubbles themselves.}

The origin of the \fb{} is still under debate. There are two main hypothesis,
which were already raised in \cite{ssf10}: active galactic nucleus (AGN)
activity, or a bipolar galactic wind. In fact, a combined origin of central accretion 
activity and  a galactic wind fed by star formation may be possible, as discussed 
in \cite{CCS13}. In that work, the authors adopted a leptonic scenario for the \fb{}
emission, whereas \cite{ca11} considered hadronic radiation. 

Numerical simulations of the dynamics of the \fb{} can be useful to unveil
relevant physical processes underlying the detected radiation. In this regard,
numerical hydrodynamical simulations of the \fb{} were performed in 2D 
\citep{gm12,gmdo12} and 3D \citep{yrrz12}. In the latter paper, the authors took
into account anisotropic diffusion of cosmic rays to explain the sharp edges of
the \fb{} in gamma rays. {In \cite{gm12},  the X-rays detected by {\it ROSAT} were
explained as thermal radiation from the surrounding medium heated by a shock, and the gamma rays from the \fb{} as 
coming from relativistic electrons in the ejecta.}

{ Specially motivated by the recent detection of a partly-ordered magnetic-field structure in the \fer{} bubble
region,}
this work presents the results of 3D magnetohydrodynamical (MHD) simulations of the
structures associated with the \fb{} in the framework of a mini-AGN event in which material is ejected from 
the central Galactic black hole \citep[see][for recent observational support for the AGN scenario]{bms13}. 
We interpret in this context 
the recent observational findings in radio, {in particular concerning the large-scale magnetic field,} 
and suggest a non-thermal nature of
the X-ray radiation detected by {\it ROSAT} that would be produced by
electrons accelerated in the shock driven by  the ejecta in
the surrounding medium. 
The GeV emission from the \fb{} could result from  
particle acceleration in the shock wave of a second explosive event in a 
recurrent-activity scenario, or
particle acceleration within the ejecta, in which shocks, turbulence, and
suitable conditions for magnetic reconnection are present. 
All through the paper, the notation $A_{b}=A/10^{b}$ has
been adopted, where $A$ has cgs units.

\section{Physical scenario} \label{phys}

We present an MHD model for the formation of the large-scale magnetic-field structure in
the \fer{} bubble region in the context of accretion-driven explosive events. The evolution and 
observational properties of the \fb{} and accompanying 
structures can be strongly influenced by the presence of this large-scale magnetic field. 

\begin{figure}
\includegraphics[width=0.35\textwidth,angle=0]{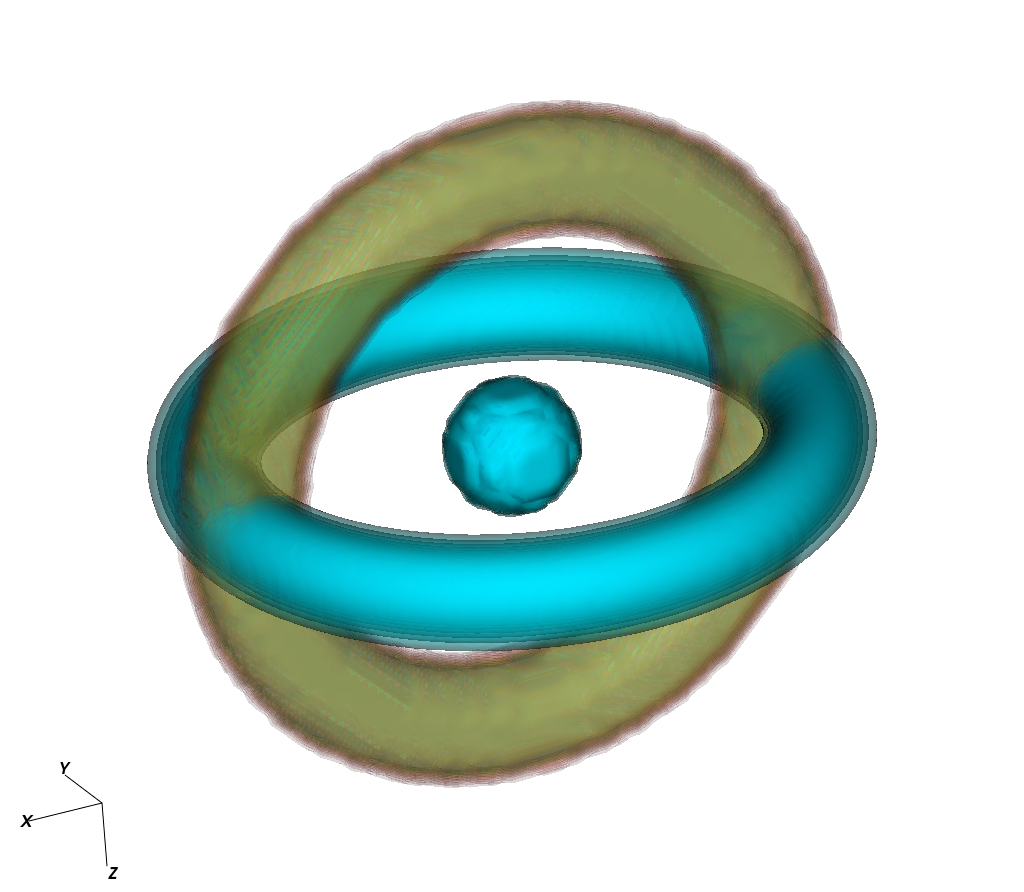}
\caption{The sketch of the initial configuration: the opaque horizontal torus is a gaseous disc with total radius $R_{\rm D}=1$~kpc and {\it small} radius 
$r_{\rm D}=0.2$~kpc; the translucent vertical ring is a magnetic
loop with the same radius and thickness as those of the
gaseous disc. The central sphere of radius $r_{\rm s}=0.2$~kpc represents a
central region with high pressure and velocity: the ejection region.}
\label{fsk}
\end{figure}

We assume that the very central region of the Galaxy is surrounded by a rotating gaseous
torus or disc with total radius  $R_{\rm D}$, effective thickness or {\it small} radius
of the torus $r_{\rm D}$, and rotation speed $\v_{\rm D}$. Loops of magnetic field
thread the central region. These magnetic loops, of strength $B_{\rm l}$, are anchored
in the disc and surround the central region as shown in Fig.~\ref{fsk}.

In the proposed scenario, the formation of the \fer{} bubble region goes as follows. A
molecular  disc, with a lifetime $\sim 10^7$~years \citep[][]{mbn11}, surrounds the
Galactic Centre. A magnetic field loop anchored in the disc also crosses the
disc polar regions. At a given time, and possibly episodically, the accretion rate in the central black
hole increases dramatically due for instance to the capture of a molecular
cloud. Such an accretion event can launch a powerful outflow in the polar
direction, as it happens in AGN. The ejecta will stretch the magnetic field
loops crossing the polar regions, and push the external medium up- and
side-wards, shocking it. Simultaneously, the external layers of the whole structure, 
where the ejecta and the external medium are in contact, will be twisted by the 
rotating gaseous disc through the surrounding magnetic field anchored to the disc and sheared 
as well by
differential rotation. As shown below, this scenario can produce the { partly-ordered magnetic-field} structure similar to that found in
S-PASS.  

\section{Simulations}\label{sim}

The simulations presented here were implemented in 3D Cartesian geometry using
the {\it PLUTO} code \citep{mbm07,mzt12}, the piece-parabolic method (PPM)
\citep{cw84,mpb05}, an HLLD Riemann Solver \citep{hlld}, and applying AMR using
the {\it Chombo} code \citep{chombo2009}. The flow was approximated as an ideal,
adiabatic gas with magnetic field, one particle species, and polytropic index of
5/3.  The adopted resolution was $48 \times 48 \times 96$ cells and 3
levels of AMR, which gives an effective resolution of $384 \times 384 \times
768$ cells. The size of the domain was $x \in [0,18 \text{ kpc}]$, $y \in [0,18
\text{ kpc}]$, and $z \in [0,36 \text{ kpc}]$. The calculations were carried out
in the Moscow State University cluster {\it Chebyshev}. The visualization of the
results were arranged with {\it VisIt}.

\subsection{Initial setup}

In the centre of our galaxy there is a massive gaseous disc of $\sim 3\times10^7\;M_{\odot}$,
a radius of 100~pc, and a rotation speed $\sim 80$~km~s$^{-1}$
\citep[see][]{mbn11}. The {simulation}
initial configuration sets an horizontal gaseous torus in the plane $XY$ { (the disc),} and a
vertical magnetic field loop in the plane $XZ$. The {torus} radius and effective
thickness are taken $R_{\rm D}=1$~kpc and $r_{\rm D}=0.2$~kpc, respectively. 
We fix the central particle density of the torus to $10^3$~cm$^{-3}$, and assume 
{ constant angular momentum} and hydrostatic equilibrium to determine the torus 
pressure and temperature distribution. { We use a truncated Keplerian gravitational potential 
$\phi\propto 1/\max(r,0.6 \rm kpc)$. Such a potential will not significantly influence the calculation results
because the speed of the shock wave at all radii is significantly larger than the escape velocity.}
The magnetic
loop radius and thickness are taken equal to those of the torus (see Fig.\ref{fsk}).
{ The initial strength of the magnetic field in the loop is 100~$\mu$G, similar to the observed value in the Galactic center
\citep{cjm10,cja11}.}
Numerical limitations have led us to adopt a larger size
of the {disc/torus} and the associated magnetic loop. 
This has also implied that the
disc is simulated rotating in the gravitational potential of the central region with a rather 
high speed, $\v_{\rm D}=800$~km/s, to have a disc angular velocity equal to the
actual angular velocity of the molecular disc in the center of the Galaxy.
Hence, in practice, the main difference between our simulation and a more realistic setup 
is the width of the inner ejection region, which is expected to have a minor impact on the 
long-term evolution of the simulated structures. 
{ In addition, the somewhat unrealistic torus properties do not allow a precise, quantitative comparison
between the simulated magnetic-field strength and the observed values. However, given that the magnetic
field is not dynamically relevant, it is determined by the fluid evolution and the obtained global field geometry 
should be reliable.} The
lifetime of the central engine activity is assumed to be much shorter than 
the disc or the dynamical scale of the simulation.

The explosive activity produces an axial outflow from a central sphere of radius $r_{\rm
s}=0.2$~kpc. The outflow, initially confined in this sphere, has been set with two different 
total energies for one simulation with one ejection:
$E_{\rm tot}=  10^{55}$ and $3\times10^{55}$~erg; a particle
density of 1~cm$^{-3}$, and a velocity of $v_{\rm e}=3.5 \times 10^{9} \mbox{ sign(z)}
E_{56}^{1/2}$~cm~s$^{-1}$ at ejection, being directed along the $Z$-axis. The
gas density of the surrounding medium is set as in \cite{bm13}:
$
n_{\rm bg}=\frac{n_0}{ \left[1+(R/R_{\rm C})^2+(z/z_{\rm C})^2 \right]^{3\beta/2} },
$
where $n_0 = 0.46 \mbox{ cm}^{-3}$, $R$ is the cylindrical radius, $z$ the distance 
from the Galactic plane, and $\beta=0.71$, $R_{\rm C} = 0.42$~kpc and $z_{\rm C} = 0.26$~pc are normalizing constants. 
The temperature of this medium is fixed to $1.2\times10^6$~K.
Provided that accretion events may be recurrent, we have performed an additional simulation with two active episodes.
An amount of energy $E_{\rm tot}=10^{55}$~erg was now twice injected, at $t=0$ and at $t=2.5$~Myr. { The central engine
activity was taken shorter than the flow crossing time, $r_{\rm s}/v_{\rm e}\sim 10^4$~yr. Being much shorter than any relevant
dynamical process in the simulation, the activity periods can indeed be considered as discrete events.}
Note that the 
formation timescale of this magnetic field structure is $\sim 100\,{\rm pc}/80$~km~s$^{-1}\approx 1$~Myr, which implies that
the magnetic loop {can keep being generated} between events.

\subsection{Results}\label{res}

 \begin{figure*}
\includegraphics[width=0.46\textwidth,angle=0]{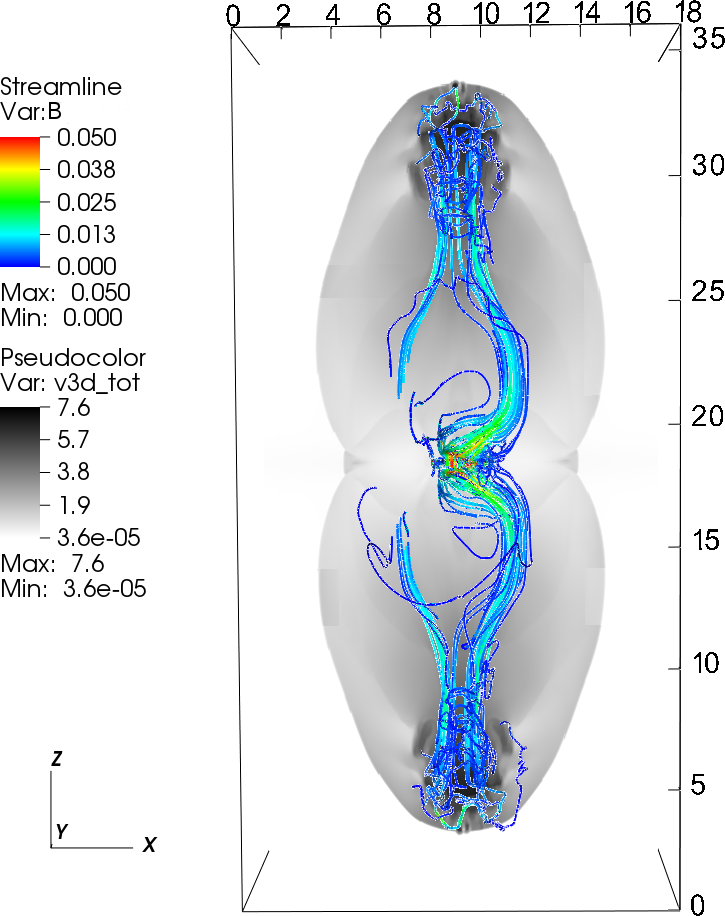}
\includegraphics[width=0.46\textwidth,angle=0]{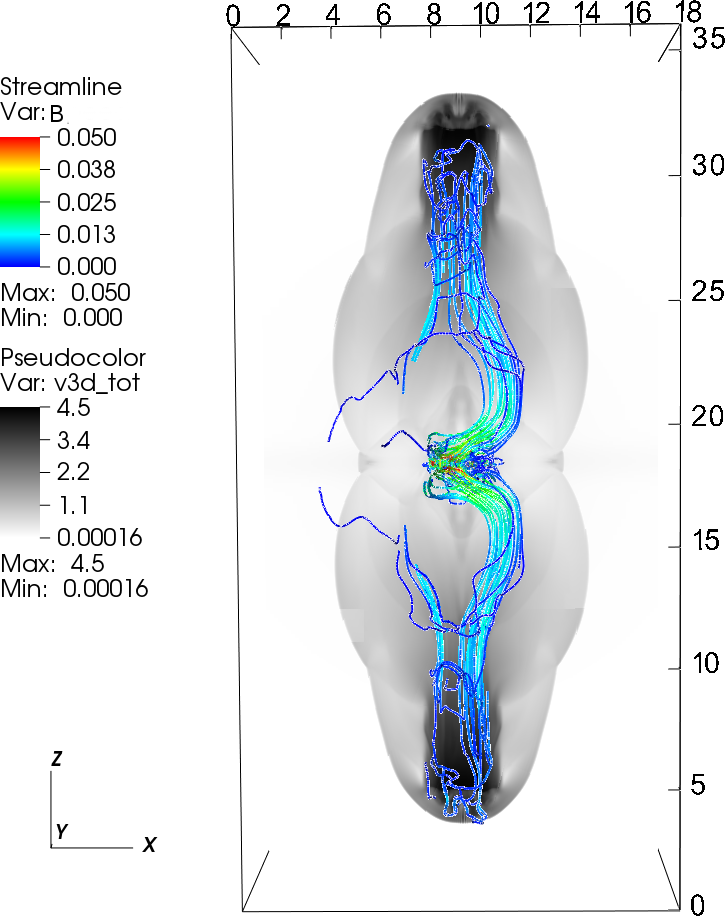}
\caption{The distribution of the gas velocity modulus and the magnetic-field lines for two
models: on the left panel, $E_{\rm tot} = 3\times10^{55}$~erg at a time
$t=2.3$~Myr; on the right panel, $E_{\rm tot} = 10^{55}$~erg at $t=3.4$~Myr. 
The velocity is presented in units of $10^8$~cm/s and the magnetic-field strength (shown by colour) in units of 400 $\mu$G.}
\label{vb}
\end{figure*}

 \begin{figure}
\includegraphics[width=0.32\textwidth,angle=0]{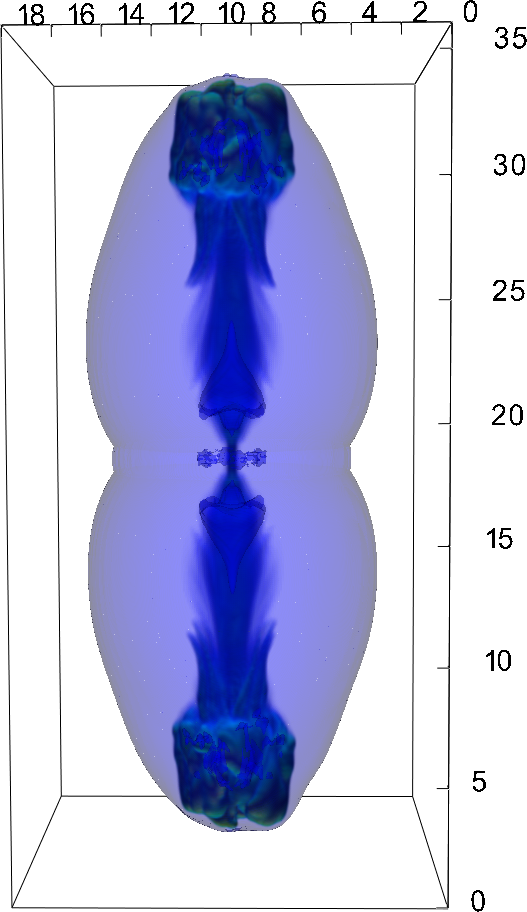}
\caption{The distribution of the ejected matter tracer (opaque blue-green)
and pressure surface (transparent magenta) indicates the position of the shock wave from the model with
$E_{\rm tot}=3\times10^{55}$~erg at $t=2.3$~Myr.}

\label{tr1}
\end{figure}

 \begin{figure}
\includegraphics[width=0.55\textwidth,angle=0]{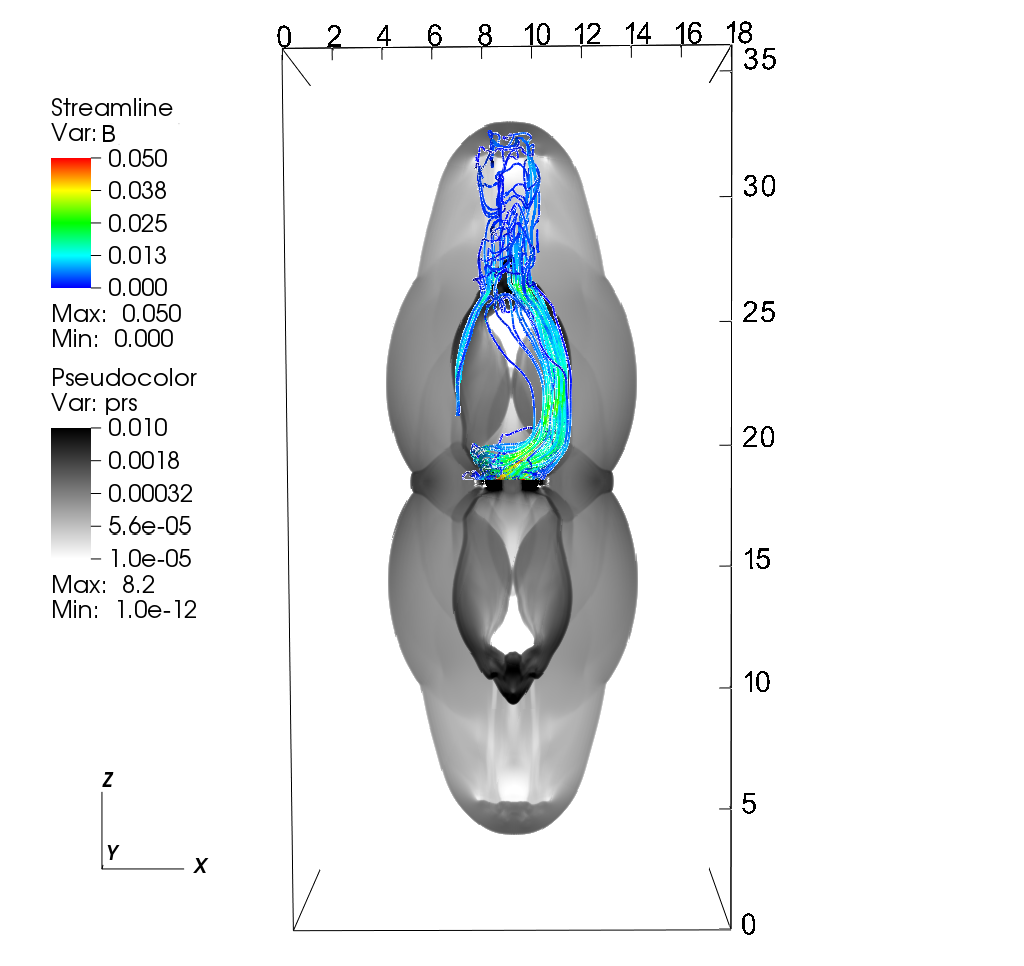}
\caption{The distribution of the gas pressure (botton and top) and magnetic-field lines (top) at $t=3.4$~Myr for a model with two active
episodes, with  
$E_{\rm tot} =10^{55}$~erg each at times $t=0$ and  $t=2.5$~Myr. The pressure is presented
in units of $1.67\times10^{-8}$~pa and the magnetic field (shown by colour) in units of 400 $\mu$G.}

\label{twoep}
\end{figure}

The distributions of velocity and magnetic-field lines are presented in
Fig.~\ref{vb} for two models, $E_{\rm tot} = 3\times10^{55}$ and $E_{\rm tot} = 10^{55}$~erg,
at the time when the size along the $Z$-direction
of the whole structure reached $\approx 15$~kpc. The expansion of the high-pressure region
blows up the bubbles and stretches the magnetic-field lines, and meanwhile the
rotation of the magnetic-line foot points twist the magnetic field into
a spiral structure. This spiral structure is similar to
that observed in \cite{CCS13}.

The ejected matter from the central engine forms a bipolar, elongated and wide
structure.  
The shock wave traveling through the Galactic halo forms an external
bubble of shocked external medium almost as wide as long.
Surrounded by this shocked external medium, the ejected
matter forms a narrower second bubble with its neck at the Galactic Centre (see
Fig.~\ref{tr1}). Covering the inner bubble, at the contact layer between the ejecta
and the shocked external medium, the pushed magnetic field is threaded, sheared 
and anchored in the rotating disc. The magnetic structure induces minor perturbations on the
contact layer.
In the simulation with two active episodes a second bubble forms inside the first one, as shown in
Fig.~\ref{twoep}. In this case, the sheared magnetic lines anchored in the molecular disc surround
the shock wave formed by the second bubble on the earlier one.
An interesting property of the computed solution is a higher speed
at the top of the bubble shocks than at their sides. This effect is produced by a decreasing background-density profile.

\section{Discussion}\label{disc}

In this work we propose an accretion-ejection, explosive scenario for the formation of the large-scale magnetic-field structure around 
the \fb{}. As shown in Fig.~\ref{tr1}, one explosion produces very extended ejecta, much larger than the \fb{} themselves, with the head
reaching up to the external boundary of the whole structure. This mismatch in size may be explained by invoking acceleration and confinement
of the GeV emitting particles only deep into the ejecta. Another possibility would come from accretion-ejection events recurrent on
timescales $\gtrsim 10^6$~Myr. Such a timescale, and the energetics derived here, would imply an average injection luminosity of $\sim
10^{41}$~erg~s$^{-1}$, compatible with cosmic ray injection.  As shown in Fig.~\ref{twoep}, the two-episodes scenario, with such a
characteristic timescale and energy budget, nicely reproduces the overall morphology of the whole \fer{} bubble region, in particular the
sharp edge of the \fb{} themselves\footnote{Such an explanation for the sharp edge of the \fb{} is only valid as long as the bubbles and the
structures at larger scales have a common origin.}. It is worth noting that, independently of the simulation, the assumption of a magnetic
structure anchored in the central molecular disc already provides a constraint on the age of the \fb{}, $\lesssim 10^7$~yr. If it were true,
such a short lifetime would disfavor the hadronic scenario. { This is so because of the long cooling time of proton-proton collisions
\citep{kab06}: $t_{\rm pp}=10^{15}/n > 10^{18} n_{-4}^{-1}\mbox{ sec } = 3\times10^{10} n_{-4}^{-1}$~yr, $\sim 10^4$ times longer than the
dynamical time of the magnetic field. This low efficiency would require a total energy only in high energy protons $\sim L_{\gamma} t_{\rm
pp}/0.17 = 2\times 10^{56}$~ergs, 10 times larger than the total energetics of the \fb{} if they are related to the large-scale
magnetic-field structure, as assumed here. Accounting for dynamical constraints, i.e. to explain the growth of the structure in the
surrounding medium on the timescales considered here, one gets an estimate on the luminosity injected into the \fb{} of $\sim 10^{41}$~ergs.
This is $\sim 10^3$ times larger than the energy required to  explain the gamma-ray luminosity, so the energy budget is not tight when
considering a leptonic origin of the gamma-ray emission.}


Let us now briefly consider a framework for the broadband detected radiation. Electrons and protons can be accelerated at the shock produced 
by the ejected matter in the surrounding medium. 
As just argued, protons are disfavored by the relatively short timescales involved in the adopted scenario. 
For electrons, on the other hand, this is not a problem, with the most efficient radiation process at low energies being synchrotron, 
and at high energies, inverse Compton (IC) with soft Galactic photons of $\sim 1$eV. The maximum energy of the photons produced via 
synchrotron from shock accelerated electrons can be estimated as 
$\epsilon_{\rm s}\approx 0.12\left(\v_{\rm s,8}\right)^2 \eta^{-1}$~keV \citep{aa99}, where $\v_{\rm s}$ is the forward-shock speed. However, 
given the slow steepening with energy above $\epsilon_{\rm s}$ of the synchrotron spectrum of shock accelerated electrons
\citep{za07}, the effective maximum energy of the synchrotron photons may be estimated about $10\times\epsilon_{\rm s}$. 
With a speed of the shock wave of $\sim 2\times10^8$~cm~s$^{-1}$, synchrotron X-rays can reach up to few keV. 
{This synchrotron radiation could explain the X-ray shell or arch found by ROSAT \citep{sef97}, which would have then a non-thermal
origin instead of a thermal one \citep[both possibilities were discussed for instance in][]{2010ApJ...724.1044S}};
IC emission from the same electrons could explain the hint of 
GeV radiation found on similar scales (see fig. 6 in \citealt{fermi12}). A diffuse lower-energy electron population, extended also beyond
the
\fb{} 
but partially embedded in the spiral-like magnetic lines, could explain the WMAP extended source. The \fb{} themselves would have an IC origin. 
They would remain confined to the fresher material of the most recent episode, or alternatively, to deeper regions of the whole region 
if only one explosion took place as it was discussed by \cite{gm12}. A hardening of the GeV radiation along the $Z$-direction has been
observed for the \fb{} \citep{hs13}. 
This hardening could be explained by the varying speed of the bubble shocks mentioned in Sect.~\ref{res}, 
affecting (at least slightly) the slope of the distribution of the GeV emitting particles. This hardening may also be consistent with a 
stronger shock at the top of the bubbles than at their sides.

\begin{acknowledgements}
We want to thank
the anonymous referee for useful and constructive comments.
The calculations were carried out in the cluster of Moscow State University {\it Chebyshev}.
We thank Andrea Mignone and the {\it PLUTO} team for the
possibility to use the {\it PLUTO} code. We also thank the {\it Chombo} team for the
possibility to use the {\it Chombo} code. We would like to thanks to {\it VisIt} visualization packet teem. 
BMV acknowledges partial  support  by  RFBR  grant  12-02-01336-a.
V.B-R. acknowledges support by DGI of the Spanish Ministerio de Econom\'{\i}a
y Competitividad (MINECO) under grants  AYA2010-21782-C03-01 and FPA2010-22056-C06-02.
V.B-R. acknowledges financial support from MINECO through a Ram\'on y
Cajal fellowship. This research has been
supported by the Marie Curie Career Integration Grant 321520.

\end{acknowledgements}

\end{document}